\documentclass[3p, times,merge]{elsarticle}
\journal{Computer Physics Communications}
\usepackage[final,notref,notcite]{showkeys}
\usepackage{amsmath}
\usepackage{amssymb}
\usepackage{bm}
\usepackage[ISO=false]{diffcoeff}[=v4]
\usepackage{listings}
\usepackage{graphicx}
\usepackage[utf8]{inputenc}
\usepackage{xcolor}
\usepackage{calc}
\usepackage{array}
\usepackage{url}
\usepackage{hyperref}

\definecolor{mydarkgray}{gray}{0.2}

\newcommand{\Hcal}{\mathcal{H}}

\newcommand{\Fcal}{\mathcal{F}}

\newcommand{\Kcal}{\mathcal{K}}

\newcommand{\vect}[1]{\boldsymbol{#1}_{\perp}}
\newcommand{\kt}{\vect{k}}

\newcommand{\bt}{\vect{b}}

\newcommand{\xt}{\vect{x}}
\newcommand{\yt}{\vect{y}}

\newcommand{\bxt}{\vect{\bar x}}
\newcommand{\byt}{\vect{\bar y}}
\newcommand{\bzt}{\vect{\bar z}}
\newcommand{\brt}{\vect{\bar r}}

\newcommand{\der}{\mathrm{d}}

\newcommand{\Tr}{\mathrm{Tr}}

\begin{document}

\begin{frontmatter}

\title{Improving the solver for the Balitsky-Kovchegov evolution equation with Automatic Differentiation}

\author[label1]{Florian Cougoulic}
\ead{florian.cougoulic@uj.edu.pl}
\author[label1]{Piotr Korcyl}
\ead{piotr.korcyl@uj.edu.pl}
\author[label1]{Tomasz Stebel}
\ead{tomasz.stebel@uj.edu.pl}
\address[label1]{Institute of Theoretical Physics, Jagiellonian University, ul. \L ojasiewicza 11, 30-348 Krak\'ow, Poland}

\begin{abstract}
The Balitsky-Kovchegov (BK) evolution equation is an equation derived from perturbative Quantum Chromodynamics that allows one to evolve with collision energy the scattering amplitude of a pair of quark and antiquark off a hadron target, called the dipole amplitude. The initial condition, being a non-perturbative object, usually has to be modeled separately. Typically, the model contains several tunable parameters that are determined by fitting to experimental data. In this contribution, we propose an implementation of the BK solver using differentiable programming. Automatic differentiation offers the possibility that the first and second derivatives of the amplitude with respect to the initial condition parameters are automatically calculated at all stages of the simulation. This fact should considerably facilitate and speed up the fitting step. Moreover, in the context of Transverse Momentum Distributions (TMD), we demonstrate that automatic differentiation can be used to obtain the first and second derivatives of the amplitude with respect to the quark-antiquark separation. These derivatives can be used to relate various TMD functions to the dipole amplitude. Our C++ code for the solver, which is available in a public repository \cite{repository}, includes the Balitsky one-loop running coupling prescription and the kinematic constraint. This version of the BK equation is widely used in the small-$x$ evolution framework.
\end{abstract}

\end{frontmatter}

\section{Introduction}

It is expected that new measurements of cross-sections in $e+\textrm{A}$ collisions conducted at the future Electron-Ion Collider \cite{Accardi:2012qut,AbdulKhalek:2021gbh} in the low-$x$ regime will shed more light on the long-sought phenomenon of gluon saturation. It would manifest itself by the fact that the rise of the cross-section with energy, predicted by the linear evolution equation derived with perturbative QCD \cite{Kuraev:1977fs,Balitsky:1978ic}, is replaced by some sort of flattening due to the non-linear effects, i.e. gluon recombination, captured by the Balitsky-Kovchegov (BK) equation
\cite{Balitsky:1995ub,Balitsky:1998ya,Kovchegov:1999yj,Kovchegov:1999ua}
or more generally the Jalilian-Marian-Iancu-McLerran-Weigert-Leonidov-Kovner (JIMWLK) equation 
\cite{Jalilian-Marian:1997jhx,Jalilian-Marian:1997ubg,Kovner:2000pt,Iancu:2000hn,Ferreiro:2001qy,Iancu:2001ad,Weigert:2000gi}. The calculation of the Deep Inelastic Scattering (DIS) cross section requires a non-perturbative model of the initial configuration of color sources in the target that is then evolved to higher energies with one of the above equations. Typically one uses the Golec-Biernat--Wusthoff (GBW) model 
\cite{Golec-Biernat:1998zce,Golec-Biernat:1999qor}
or McLerran-Venugopalan (MV) model 
\cite{McLerran:1993ni,McLerran:1993ka,McLerran:1994vd} to obtain the initial distribution of color charges. In either case, the values of the free, non-perturbative parameters of these models 
must be determined by comparison with experimental data. In the most naive approach, one constructs an appropriate $\chi^2$ function that is then minimized numerically with respect to the model parameters. Such a numerical minimization often requires the evaluation of gradients in the parameter space of the $\chi^2$ function in order to approach the minimum. 
The derivatives can only be accessed if the estimates of $\chi^2$ are available for at least two different values of the initial parameters, {i.e.}~for each parameter an additional solution of the evolution equation is needed. This procedure must be performed at each step of the iterative evolution. 

In this contribution, we observe that one can supplement the simulation with an automatic computation of derivatives through Automatic Differentiation. In this way, the solution of the evolution equation would provide the analytical value of the derivatives with respect to all parameters of the initial condition at any time at a reasonable additional cost. 
Automatic Differentiation is a programming technique in which real numbers are lifted up to dual numbers: objects containing the value and all its derivatives up to a given order. In our implementation, we provide an example with two parameters and we compute all results together with the first and second derivatives (full Hessian matrices) with respect to these two parameters. The use of our software should facilitate the investigation of more sophisticated initial condition models, which may have more tunable parameters. Such a situation already occurs in the preparatory work, where saturation signals are sought in DIS experiments on heavy ions, and the parameters of the initial condition may have to be adapted for each type of ion used.

Although there exist automatic tools that, given a simulation source code, generate a new version where the calculation of derivatives is automatically implemented \cite{autodiff}, we provide our own implementation. This is motivated, first of all, by the fact that we are interested in the first derivatives and the Hessian matrix of only a few specific parameters that may appear independently in different primitives of the program. 
Our second motivation is to automatically estimate the first and second derivatives with respect to the dipole size $\brt$ of the dipole amplitude. However, the BK equation is an integrodifferential equation for the dipole amplitude $S_{\brt = \bxt - \byt}(\eta)$ which contains a convolution with respect to the daughter dipoles' sizes $\boldsymbol{r}_{xz}$ and $\boldsymbol{r}_{zy}$ where $\brt = \boldsymbol{r}_{xz} + \boldsymbol{r}_{zy}$. 
Thus, if we choose one daughter dipole size $\boldsymbol{r}_{xz}$ as the independent integration variable, the dependence of the other daughter dipole size $\boldsymbol{r}_{zy} = \boldsymbol{r}_{zy}(\brt)$ becomes non-trivial. Therefore, it is not straightforward and automatic to track down the differentiable variable $\brt$ in the various places of the source code and we cannot rely on automatic tools.

In the following, we provide more details about the leading-order (LO) BK equation and the more advanced form that includes several phenomenologically relevant improvements that we used in our work. We also provide a brief discussion of the convergence of the BK equation and its derivatives with respect to dipole size. In Section \ref{sec. description}, we discuss the changes necessary to implement Automatic Differentiation into the framework that solves the BK equation. In Section \ref{sec. results}, we show the comparison of our results against the existing software. We take this opportunity to provide some remarks on the performance. Section \ref{sec. applications} is devoted to the applications. First, we show the quality of the derivatives of the DIS cross-section with respect to the initial condition parameters. Second, we present functions 
$\mathcal{K}_+$ and $\mathcal{K}_-$ 
needed for the calculation of the TMD distributions in the Gaussian approximation. These functions are constructed from derivatives with respect to the quark-antiquark separation of the dipole amplitude.
By comparing the results obtained with automatic differentiation against the divided differences method applied directly to the dipole amplitude, we provide evidence for the superiority of the AD approach. Eventually, we give conclusions and outlook in Section \ref{sec. conclusions}.

The code described in this work is publicly available in the source code repository \cite{repository}: {\url{https://bitbucket.org/piotrekkorcyl/bk_with_ad/src/main/}}. 

\section{Balitsky-Kovchegov evolution equation}

The LO BK equation reads \cite{Balitsky:1995ub,Balitsky:1998ya,Kovchegov:1999yj,Kovchegov:1999ua}
\begin{equation}
\frac{\partial S_{\bxt \byt}(\eta)}{\partial \eta} = \frac{\bar{\alpha}_s}{2 \pi} \int d^2 \bzt\   \mathcal{M}_{\bxt\,\byt\,\bzt} \big[ S_{\bxt\bzt}(\eta) S_{\bzt\byt}(\eta) - S_{\bxt\byt}(\eta)\big],
\label{eq. BK simple}
\end{equation}
where $\bar{\alpha}_s= \frac{\alpha_s N_c}{\pi}$, and $\alpha_s$ is the strong coupling constant of QCD taken here as a constant, $\eta$ is the rapidity variable, and $S_{\bxt \byt}(\eta)$ is the dipole amplitude describing the amplitude of scattering the pair of quark and anti-quark of size $\brt = \bxt - \byt$ on a proton in the limit of very high energies.
The dipole kernel reads
\begin{equation}\label{eq:DipoleKernel}
    \mathcal{M}_{\bxt\,\byt\,\bzt} = \frac{(\bxt - \byt)^2}{(\bxt - \bzt)^2 (\bzt - \byt)^2}.
\end{equation}
In the following, we assume translation and rotational invariance (which also implies the assumption of an infinite target size)\footnote{Approaches to BK going beyond assumptions of infinite target size were also developed, see for example \cite{Berger:2010sh,Rezaeian:2013tka,Cepila:2018faq}, but these are beyond scope of this manuscript. } so the dipole amplitude cannot depend on the orientation of the vector $\brt$ but only on its length $r=|\brt|$, $S(r,\eta)$. The most convenient way to estimate the integral over $\bzt$ is to introduce radial coordinates, $r_z = |\bxt - \bzt|$ and $\phi = \angle (\bzt, \bxt)$. From assumed symmetries, the origin of the coordinate system can be set at $\bxt$ without any loss of generality.
 Hence, rewritten in radial variables Eq.~\eqref{eq. BK simple} reads 
\begin{equation}
\frac{\partial S(r,\eta)}{\partial \eta} = \frac{\bar{\alpha}_s}{2 \pi} \int d\phi \, dr_z\, r_z \, \frac{ r^2}{r_z^2 (r^2+r_z^2-2rr_z\cos\phi)}  \left[ S(r_z,\eta) \, S\left( \sqrt{r^2+r_z^2-2rr_z\cos\phi},\eta \right) - S(r,\eta)\right].
\label{eq. BK simple radial}
\end{equation}

Although for the discussion of Automatic Differentiation in this context the form of the LO BK equation given by Eq.~\eqref{eq. BK simple} is sufficient, the numerical results that we present in Sections \ref{sec. results} and \ref{sec. applications} were obtained using the form of the BK equation derived in Ref.~\cite{Ducloue:2019ezk} which takes into account several additional physical effects, such as the running of the coupling constant with the energy scale, resummation of subleading corrections \cite{Andersson:1995ju,Kwiecinski:1996td, Motyka:2009gi}, etc. The discussion of the origin of these improvements is out of scope of the present work, however, for completeness, we quote the full form of the equation that is implemented in our software,
\begin{multline}
\frac{\partial S(r,\eta)}{\partial \eta} = \int d\phi \, dr_z\, r_z \,
\left[ \frac{\bar{\alpha}_s(r)}{2 \pi r^2_{z}} \left( \frac{r^2}{r^2_{zy} +\epsilon^2} + \frac{\bar{\alpha}_s(r_{z})}{\bar{\alpha}_s(r_{zy})}-1 + \frac{r^2_{z}}{r^2_{zy}+\epsilon^2} \left(\frac{\bar{\alpha}_s(r_{zy})}{\bar{\alpha}_s(r_{z})}-1 \right) \right) \right] 
\times \\
\times \big[ S(r_z, \eta - \delta_{r_z; r}) S(r_{zy}, \eta - \delta_{r_{zy}; r}) - S(r, \eta)\big],
\label{eq. BK improved}
\end{multline}
where $r_{zy} = \sqrt{r^2+r_z^2-2r r_z\cos\phi}$. The shifts in $\eta$ in the dipole amplitudes are given by $\delta_{r_z; r} = \max\Big\{0, 2 \log \frac{r}{r_{z}} \Big\}$ and similarly $\delta_{r_{zy}; r} = \max\Big\{0, 2 \log \frac{r}{r_{zy}} \Big\}$. 
Whenever the rapidity argument of the dipole amplitude is negative in the above square bracket, one uses the value of the dipole amplitude at the initial condition \cite{Ducloue:2019jmy}.

For $\alpha_s(r)$ we take the following prescription \cite{Lappi:2013zma},
\begin{equation}
    \alpha_s(r) = \theta(r-r_{\textrm{max}}) \alpha_s^{\textrm{freezed}} + \theta(r_{\textrm{max}}-r) \Big(\big(2 b_0 \log(\frac{2 C}{r \Lambda}) \big)^{-1} + \epsilon_{\alpha_s} \Big),
    \label{eq. alpha_s}
\end{equation}
with $\Lambda = 0.241$, $\alpha_s^{\textrm{freezed}} = 0.7$ and $r_{\textrm{max}}$ defined as
\begin{equation}
    r_{\textrm{max}} = \frac{2 C}{ \Lambda} \exp\Big(-0.5 \big( \alpha_s^{\textrm{freezed}} b_0 \big)^{-1} \Big).
    \label{rmax_def}
\end{equation}
$C$ is a tunable parameter and has been set to $3.8078$ \cite{Lappi:2013zma}, while $b_0$ is the zeroth coefficient of the $\beta$-function in QCD, $b_0 = (11 N_c-2 N_f)/(12 \pi) \approx 0.716$, where number of colors $N_c=3$ and number of quark flavours $N_f=3$.  Although $r$ and $r_z$ are positive, it may happen that $r_{zy}$ is zero and therefore we introduced $\epsilon_{\alpha_s}$ in Eq.~\eqref{eq. alpha_s} as a regularization and we set it to $10^{-7}$.

\subsection{Numeric solution of the BK equation}

The typical approach to solving numerically Eq.~\eqref{eq. BK simple} or Eq.~\eqref{eq. BK improved} is to discretize the rapidity variable $\eta$ and introduce a discrete grid in the transverse plane for the $\bzt$ vector. In radial coordinates this amounts to discretizing the radial variable $r_z$ and the angle $\phi$. It is known from previous studies \cite{KC5_Motyka:2009gi} that the most suitable radial variable for the evaluation of the convolution in Eq.~\eqref{eq. BK simple} is the logarithmic variable. Then, starting with a given analytic model for the initial condition $S(r, \eta=0)$, one uses the Euler scheme to advance the dipole amplitude $S(r, \eta)$ by a small increment $\delta_{\eta}$ in $\eta$. This requires the evaluation of the convolution with respect to $r_z$ and $\phi$, where usually $r_z$ is discretized on a logarithmic scale. In our implementation, we used the Simpson $\frac{3}{8}$ rule for both the radial and angular variables. The missing data for the dipole amplitude was obtained by linear interpolation, while the kernel and the running coupling constant were evaluated analytically.

\subsection{BK equation for the first and second derivative of dipole amplitude}

The BK equation is free of ultraviolet divergences, this follows from the probability conservation of the wave function where divergences of \textit{virtual} diagrams cancel divergences of \textit{real} diagrams when the emitted gluon $\bzt$ is at the location of one of the two parents $\bxt$ or $\byt$.
Thus, the apparently singular behavior of the dipole kernel Eq.~\eqref{eq:DipoleKernel} when $\bzt \rightarrow \bxt$ or $\bzt \rightarrow \byt$, is compensated by the combination of dipoles distributions $[S_{\bxt\bzt}S_{\bzt\byt}-S_{\bxt\byt}]$ in the integrand of Eq.~\eqref{eq. BK simple}.
As an illustration, for a Gaussian initial condition of the dipole distribution, the combination of dipoles $[S_{\bxt\bzt}S_{\bzt\byt}-S_{\bxt\byt}]$ expanded around small separation $\bzt-\bxt$ scales as $\mathcal{O}\left((\bzt-\bxt)^2\right)$, and the integrand in Eq.~\eqref{eq. BK simple} is free of divergence at said point.
This conclusion for UV-finiteness is unchanged when derivatives are taken {w.r.t.} the parent dipole size $\brt$, but due to the Leibniz rule individual terms may appear singular.

The regulator $\epsilon = 5\cdot 10^{-4}\, \text{GeV}^{-1}$ is introduced in (\ref{eq. BK improved}) to improve the numerical stability of the equations. 
We emphasize, that this regulator does not regulate the UV behavior of the equations, since they are already UV finite by themselves. In particular, $\epsilon$ can be smaller than the intrinsic UV scale provided by the lattice spacing. 
We checked that varying $\epsilon$ in a wide range from $5\cdot 10^{-5}\, \text{GeV}^{-1}$ to $5\cdot 10^{-3} \, \text{GeV}^{-1}$ had negligible impact on the results.
On the other hand, $\epsilon$ stabilizes the evaluation of the derivatives of the dipole amplitude {w.r.t.} $r$ since no $0/0$ symbol is encounter in the limit $r_z \to r$, $\phi \to 0 $.
Note that the Jacobian for the polar coordinate provides $r_z$ that improves the behavior of the integrand when $r_z \rightarrow 0$. Hence, no regulator was needed there to evaluate the dipole amplitude and its first two derivatives {w.r.t.} $r$.

\section{Description of implementation details}
\label{sec. description}

Automatic differentiation (AD) \cite{ad_lecture_notes, ad_book} has a long history in the computer science literature. It provides a convenient way of estimating analytical derivatives of computer programs.

\subsection{Forward-mode AD and dual numbers}

In the present application, we employ a forward-mode implementation of AD because of its better performance for a small number of input parameters compared to reverse-mode \cite{BARTHOLOMEWBIGGS2000171,Margossian_2019}. 

In the case of first derivatives, a real number is promoted to the abstract object called {\it dual number}, which is represented as an ordered pair of real numbers $(x,x')$. The first component $x$ follows ordinary arithmetics, whereas rules for $x'$ can be easily deduced from calculus on infinitesimal numbers. For example, sum, multiplication, $\sin$ function on dual numbers are given by: 
\begin{align}
    (x, x') + (y, y') &:= (x+y , x'+y'), \nonumber \\
    (x, x') \cdot (y, y') &:= (xy , xy'+ y x'), \nonumber\\
    \sin(x, x') &:= (\sin(x) , x'\cos(x) ). \nonumber
\end{align}

Such rules can be defined for all elementary operations appearing in the algorithm. Second derivatives can be calculated in AD by defining dual numbers as triples of real numbers $(x, x', x'')$, where the arithmetic of the third component follows from the expansion of infinitesimal numbers up to the second order.

In our code, we replace real numbers with dual numbers defined as a C++ class as shown in Listing \ref{list. dual class} in \ref{sec. implementation}. All necessary numeric primitives needed for the calculation are rewritten to operate on dual numbers and are contained in a single header file. We use operator overloading in C++ to handle dual numbers in order to increase the readability of the original source code, see Listing \ref{list. mult operator} in \ref{sec. implementation} for the example of a multiplication operator.

\subsection{Storage of the dipole amplitude}

The values of the dipole amplitude are stored in an array for each value of the radial variable $r$. We note that appropriate \verb|get| and \verb|set| functions had to be implemented in order to ensure the differential chain rule. The \verb|get| function always returns the result of a linear interpolation since the argument may not correspond to a node in the grid. This will often be the case in the term $ S\left( \sqrt{r^2+r_z^2-2rr_z\cos\phi},\eta \right)$ in Eq.~\eqref{eq. BK simple radial}. In this particular case, since the argument depends itself on the variable $r$ over which derivatives are evaluated, one needs to supplement the result of the interpolation with the additional term $\partial \Big( \sqrt{r^2+r_z^2-2rr_z\cos\phi} \Big) / \partial r$ to implement the differential chain rule at the level of the array,
\begin{equation}
    \Big( S \big( f(r) \big) \Big)' \simeq  \ \Big\{ S'\big[ i^+ \big ] + (S'\big[ i^+ \big ] - S'\big[ i^- \big ]) * \big( f(r) - i^+ \big) \Big\} f'(r),
\end{equation}
where the integers $i^+$ and $i^-$ are the nearest larger and smaller grid nodes respectively and $S'\big[ i^+ \big ]$ and $S'\big[ i^- \big ]$ their corresponding values of the derivative of the amplitude with respect to $r$. The derivatives $S'\big[ i^+ \big ]$ and $S'\big[ i^- \big ]$ can be directly accessed from the array, however, the term $f'(r)$ must be inferred from the derivative of the argument. An adequate formula for the second derivative has also been implemented. In the case of the improved BK equation Eq.~\eqref{eq. BK improved} the dipole amplitude has to be stored in a two-dimensional array for all values of rapidity $\eta$ and separation $r$. Since the rapidity shift also may depend on the variable $r$, a generalized pair of \verb|get| and \verb|set| functions were implemented to handle the chain rule in the case of two arguments depending on $r$.

In order to ensure the differentiability of the entire algorithm we replaced all non-differentiable functions by their differentiable counterparts. In particular, instead of the $\max$ function defining the shift in the rapidity of the dipole amplitudes in Eq.~\eqref{eq. BK improved} we used the expression shown in Eq.~\eqref{eq. max},
\begin{equation}
        \max(0,y) \equiv \frac{1}{2} \Big( y + \sqrt{y^2 + 0.0001} \Big). 
\label{eq. max}
\end{equation}

Similar changes have been incorporated into the definition of the running coupling constant which for phenomenological reasons needs to be set to a constant value for distances larger than $r_{\textrm{max}}$. The change from the running behavior to a constant one was implemented using a differentiable version of the Heaviside $\theta$ function.

\subsection{Multi-threading}

In order to decrease the run time, we parallelize the calculations using multi-threading \verb|openMP| library. The update of the dipole amplitude for each separation $r_{xy}$ involves the evaluation of the convolution and can be performed and stored independently. Hence, the parallelization does not involve any critical sections and can be easily implemented with an appropriate pragma.

\section{Results}
\label{sec. results}

In this Section we provide a discussion of the comparison of the results obtained with our software with the implementation used in Ref.~\cite{Caucal:2023fsf} together with some remarks on the overall performance and overhead generated by the use of dual numbers.

\subsection{Comparison with existing software}
\label{sec. comparison}

As a test of our software, we reproduced the evaluation of the dipole amplitude $S(r, \eta)$ with the results obtained with the software used in Ref.~\cite{Caucal:2023fsf}. We start with the initial condition given by the MV model \cite{McLerran:1993ni,McLerran:1993ka}
\begin{equation}
    S(r, \eta=0) = e^{-r^2 Q_0^2 \log \left(\tfrac{1}{r \Lambda} + e\right) },
    \label{MV_init_cond}
\end{equation}
with $e$ being the Euler constant.
The initial values of the parameters are $Q_0^2=6\times 0.104 \, \text{GeV}^2$ and $\Lambda=0.241\, \text{GeV}$ that correspond to the case of a heavy nucleus. The value $0.104 \, \text{GeV}^2$ was obtained from the fit of the MV model to the electron-proton scattering data in \cite{Lappi:2013zma}, we multiply it by a factor of 6 which represents the factor $A^{1/3}$ for a heavy nucleus with the $A=216$, close to the lead mass number. In phenomenological calculations, one would modify scaling $A^{1/3}$ by multiplying it by a constant $c<1$, so that the nuclear effects are not overestimated \cite{Kowalski:2007rw}. Later, we will promote $\Lambda$ and $Q_0$ parameters to dual numbers and calculate the first and second derivatives of the dipole amplitude $S(r,\eta)$ with respect to these parameters.

\begin{figure}
\begin{center}
\includegraphics[width=0.495\textwidth]{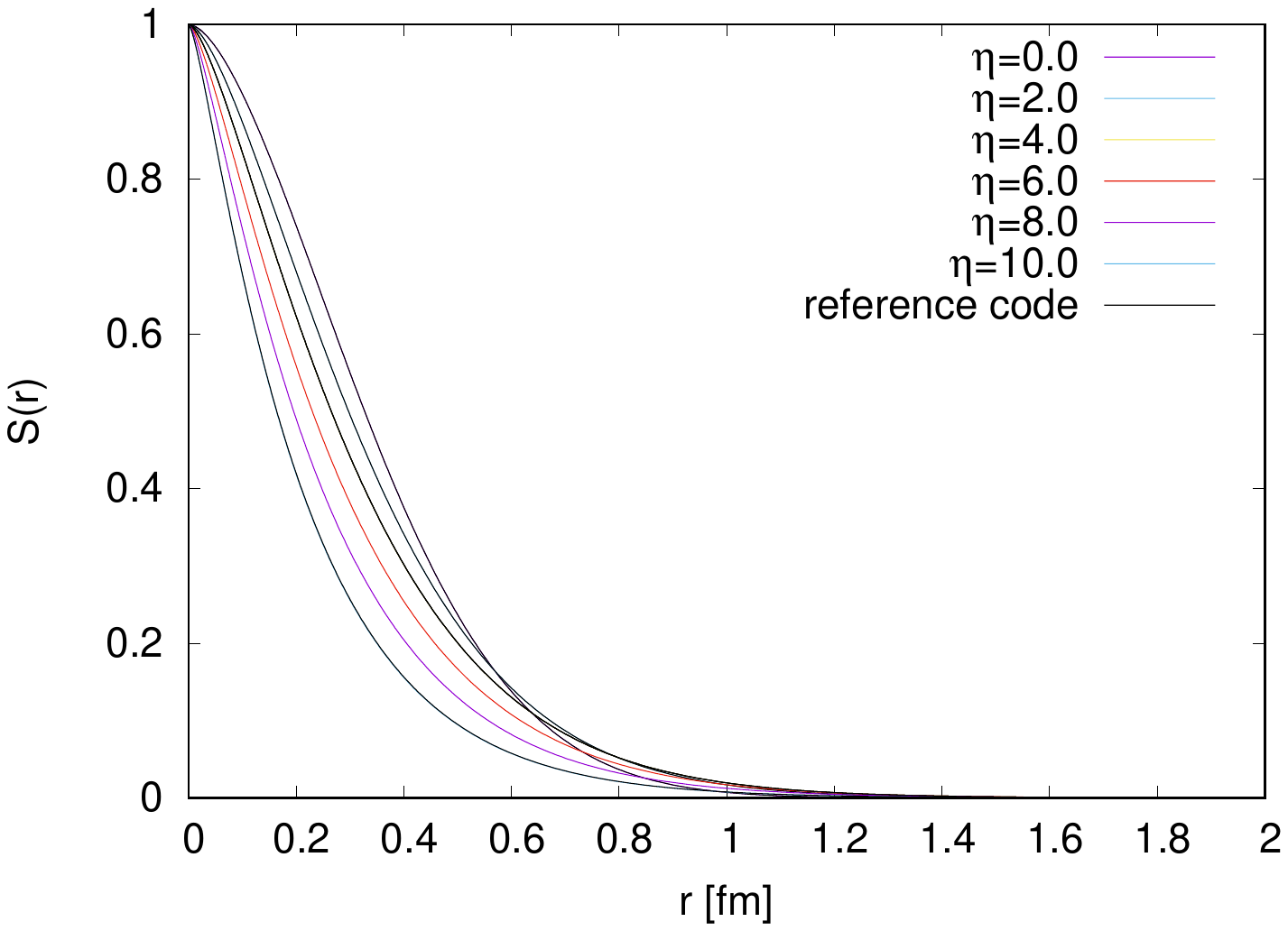}
\includegraphics[width=0.495\textwidth]{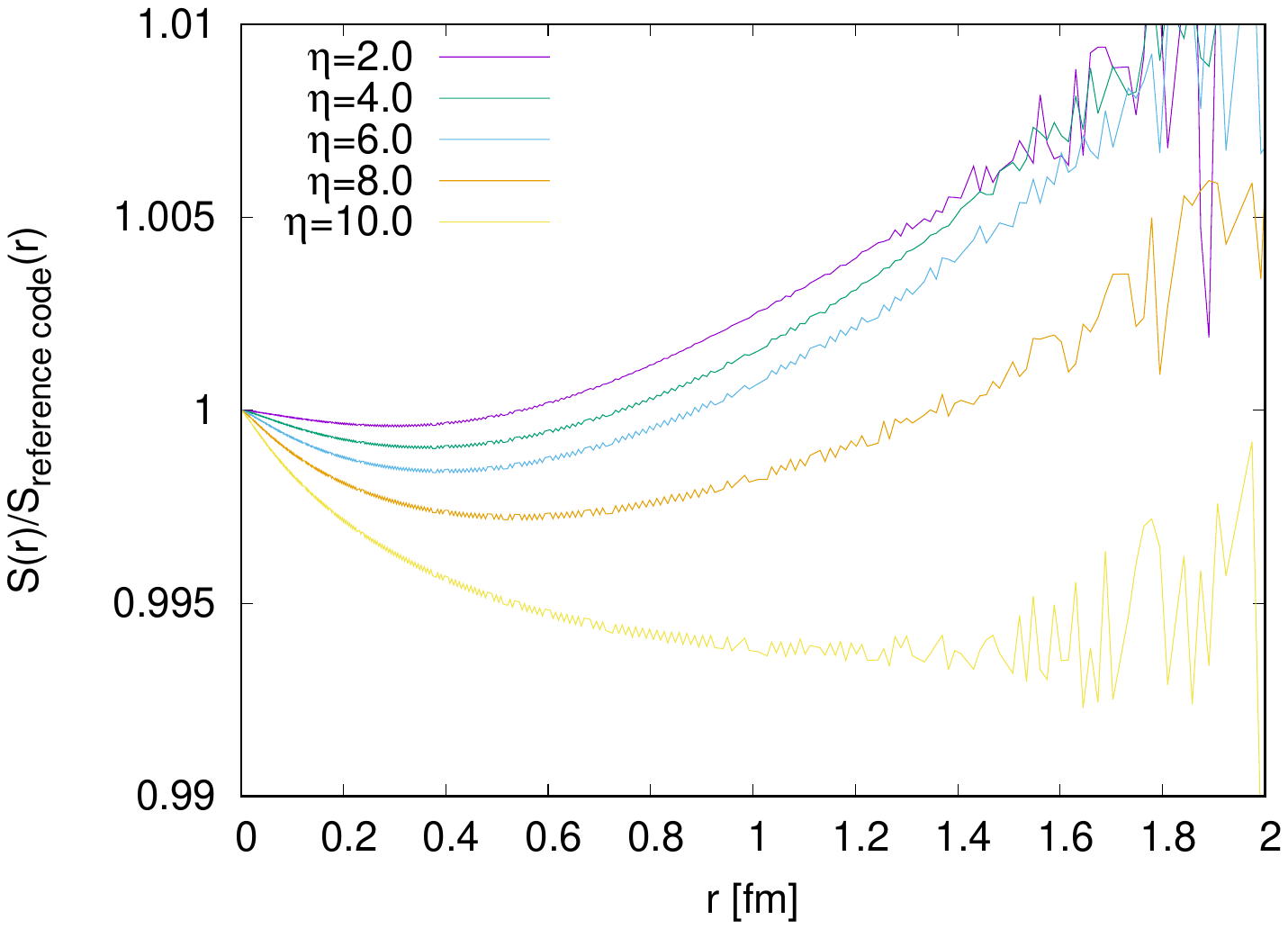}
\caption{Comparison with software from Ref.~\cite{Caucal:2023fsf}. The left panel shows the dipole amplitude obtained with our software plotted in color and obtained using reference software in black. The lines overlap hiding one beneath the other. In the right panel, we show their ratios for different values of $\eta$. \label{fig. comparison}}
\end{center}
\end{figure}

We show the comparison in Fig.\ref{fig. comparison}. On the left panel, we plot $S(r,\eta)$ for a wide range of $\eta$. Black lines correspond to the results from the reference code \cite{Caucal:2023fsf} and color lines from our implementation. Additionally, on the right panel, we show the ratio of the dipole amplitudes to demonstrate the behavior of the results for large distances. Both codes have been run with 2000 points for the radial variable $r_z$ and 80 points for the $\phi$ variable. The minimal and maximal range of $r$ have been chosen accordingly in both runs. We find very good agreement between our solver and the reference code at a permille level.  

\subsection{Comparison with divided differences}

\begin{figure}
\begin{center}
\includegraphics[width=0.495\textwidth]{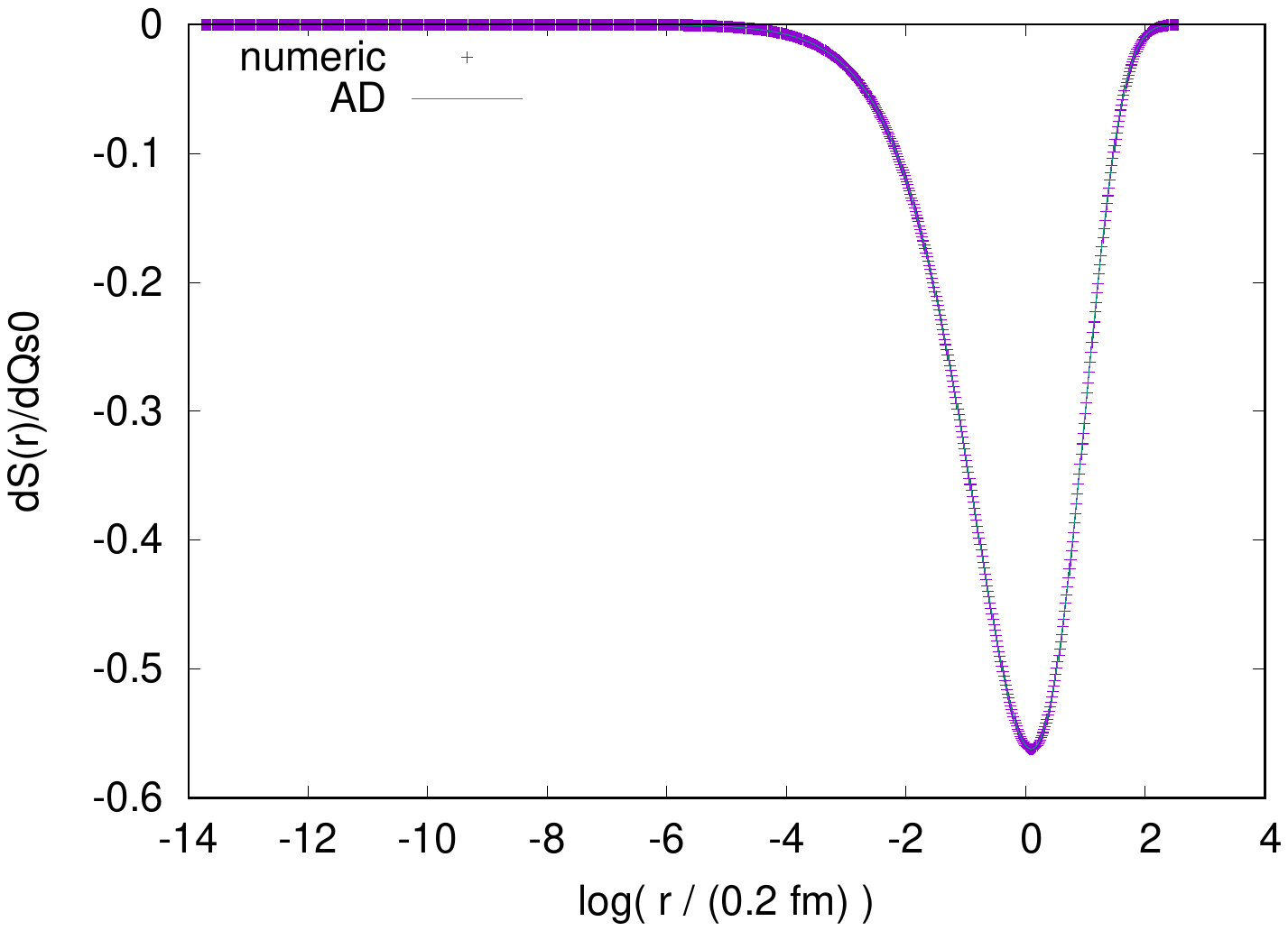}
\includegraphics[width=0.495\textwidth]{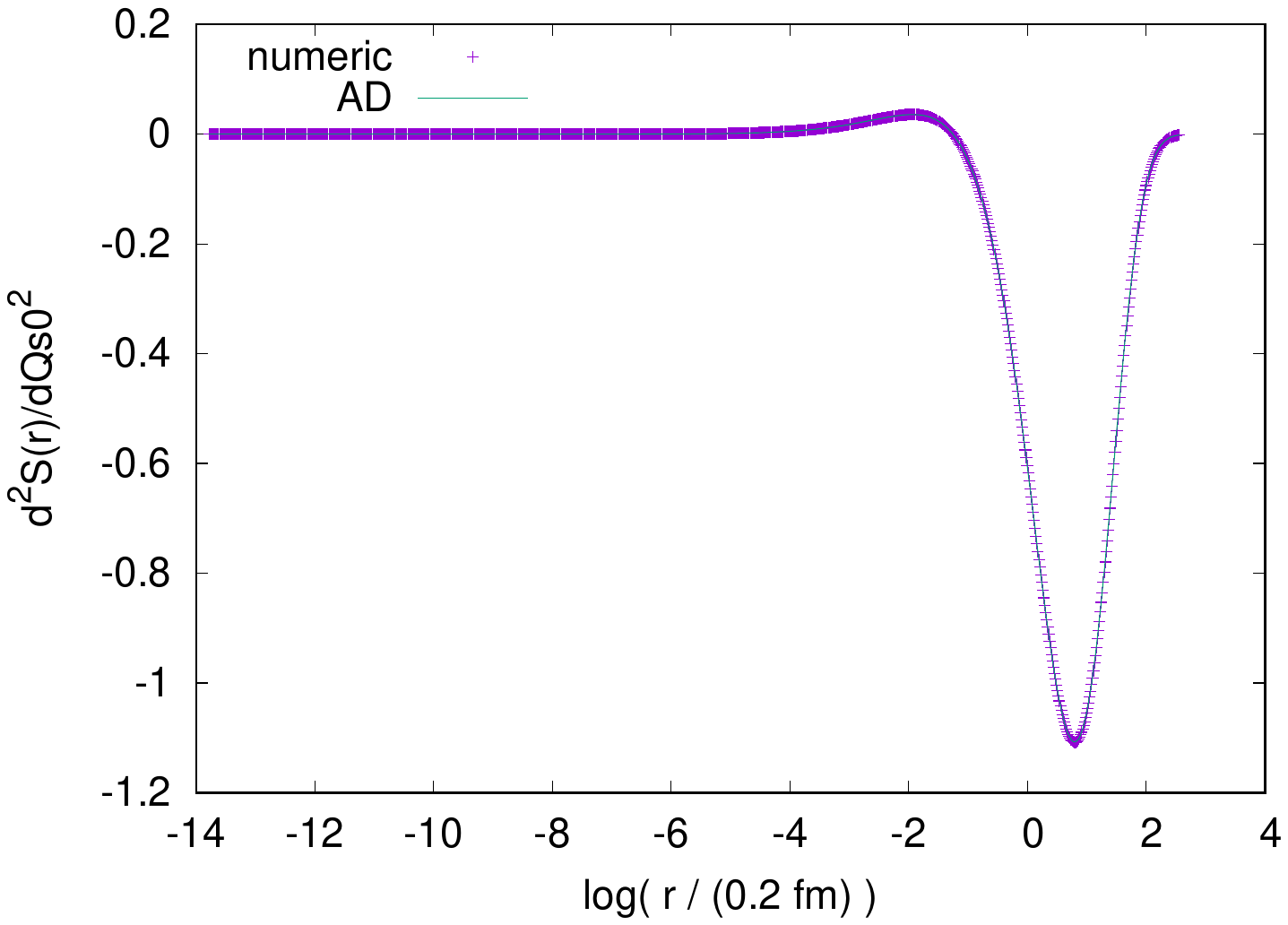}
\caption{First and second derivative with respect to the $Q_0$ parameter of the dipole amplitude evolved up to $\eta=10$ as a function of $r$. The solid line corresponds to the result obtained using AD, while the data points were calculated by divided differences (\ref{first_der_Q0},\ref{second_der_Q0}) by varying $Q_0$ by about 6\% by carrying two additional simulations. \label{fig. qs0 derivative}}
\end{center}
\end{figure}

In order to demonstrate the calculation of derivatives with respect to the parameters of the initial condition we plot in Fig.\ref{fig. qs0 derivative} the comparison of the first (left panel) and second derivative (right panel) against the estimate from divided differences of the dipole amplitude $S(r, \eta=10)$ as a function of $r$. The target value of $Q_0=0.789937 \, \text{GeV}$ and the divided difference were obtained using two additional simulations at $Q^+_0=0.839937\, \text{GeV}$ and $Q^-_0=0.739937\, \text{GeV}$. The first derivative was estimated using a symmetric difference
\begin{equation}
    \frac{d S(r,Q_0)}{d Q_0} \approx \frac{S(r,Q^+_0) - S(r,Q^-_0)}{Q^+_0 - Q^-_0},
    \label{first_der_Q0}
\end{equation}
while the second derivative was estimated according to
\begin{equation}
    \frac{d^2 S(r,Q_0)}{d^2 Q_0} \approx \frac{S(r,Q^+_0) + S(r,Q^-_0) - 2 S(r,Q_0)}{\big( \frac{1}{2} (Q^+_0 - Q^-_0) \big)^2}.
    \label{second_der_Q0}
\end{equation}
We see perfect agreement between the two estimates in the entire range of $r$.

\subsection{Performance}
\label{sec. performance}

In this Section, we comment on the performance of our implementation. In particular, we provide timings from which the overhead generated by automatic differentiation can be estimated. We repeated our experiments for three native data types with different precision and gathered the results in Table \ref{tab. performance}. We benchmarked our implementation on a two-socket AMD EPYC 7F52 16-Core Processor with hyper-threading running with 64 openMP threads. Since the dipole amplitude rapidly vanishes with increasing $r$ the usage of the float type is not recommended as it quickly generates invalid results. 

The upper part of Table \ref{tab. performance} shows the scaling with the number of discretized points of the radial variable $r$ (we name it "size"). The discretization of angular variable $\phi$ is fixed to 80 points for all the runs. The third column shows the time in seconds of 10 steps of the evolution (that correspond to $\eta=0.5$) when no AD is used. The scaling is quadratic (increasing the size by a factor of 2 increases the time fourfold) since for each $r$ value one needs to sum contributions from the whole grid. The run in long double precision (16 bytes) is the most stable, however, it takes approximately four times more time than the run in double precision (8 bytes). 
The fourth column shows similar timings for the simulation together with the calculation of first and second derivatives with respect to the dipole size $r$. In principle, one can obtain the $\partial S(r) / \partial r$ and $\partial^2 S(r) / \partial r^2$ from the dipole amplitude itself, using the divided differences approach. However, this method fails quickly for larger values of $r$ as we demonstrate later in the text. Calculating these derivatives using AD increases the time by a factor of 2. This can be compared with the observation that AD in fact solves simultaneously three coupled equations: for $S(r,\eta)$, $S'(r,\eta)$ and $S''(r,\eta)$, hence the expected increase factor would be 3. One possible source of improved efficiency is the reuse of data: see for example Listing \ref{list. exp function} for the wrapper of the \verb|exp| function where only one call to the library \verb|exp| function is necessary. The fifth column contains timings when we additionally evaluate first and second derivatives with respect to two initial condition parameters including the $2\times2$ hessian matrix. 
In the naive approximation of the Hessian matrix by the divided differences for the case of two parameters, one would need to evaluate the function 6 times: the function itself, plus and minus the increment in each parameter to get the central differences for the first derivatives and the second derivatives and one additional evaluation with both parameters incremented for the off-diagonal entry of the hessian. Counting also supplemental calculations needed for the derivatives over $r$ yields an expected factor 8 with respect to the cost without AD. Yet, we observe only a four-fold increase in time between columns 3 and 5. 

The scaling with the number of parameters over which we calculate derivatives is shown in the second part of Table \ref{tab. performance}. The scaling is roughly quadratic with the number of additional differentiable parameters, which could have been expected since the cost is dominated by the calculation of the Hessian matrix which scales quadratically with the number of parameters. We keep the derivatives over $r$ explicitly separated since they do not contribute to the Hessian matrix. 

\begin{table}
    \begin{center}
    \begin{tabular}{|c|c|c|c|c|}
    \hline
    data type & size & without AD & $r$ derivatives only & $r$ derivatives + Hessian(2 params)\\ 
    \hline 
    \verb|double| & 250 & 0m3(1) & 0m5(0.5) & 0m12(1)\\ 
    \verb|double| & 500 & 0m11(1) & 0m22(1) & 0m50(1) \\ 
    \verb|double| & 1000 & 0m46(1) & 1m29(1) & 3m22(2)\\ 
    \verb|long double|  & 1000 & 3m52(3)  & 6m11(2) & 13m51(6)\\ 
    \verb|double| & 2000 & 3m3(1) & 5m50(4)  & 13m47(3) \\ 
    \hline
    \hline
    data type & size & mode& time &   --- \\ 
    \hline
    \verb|double| & 1000 & without AD & 0m46(1) &\\
    \verb|double| & 1000 &  $r$ derivatives only & 1m29(1) &\\
    \verb|double| & 1000 &  $r$ derivatives + Hessian(1 param) & 2m12(1) &\\ 
    \verb|double| & 1000 & $r$ derivatives + Hessian(2 params) & 3m22(2) & \\ 
    \verb|double| & 1000 &  $r$ derivatives + Hessian(3 params) & 5m4(2) &\\ 
    \verb|double| & 1000 &  $r$ derivatives + Hessian(4 params) & 7m20(5) &\\ 
    \verb|double| & 1000 &  $r$ derivatives + Hessian(5 params) & 9m42(3) & \\
    \hline
    \end{tabular}
    \caption{Comparison of run times at different level of precision with and without Automatic Differentiation. Shown are timings for 10 steps of the evolution (corresponding to maximal rapidity of $\eta=0.5$) on a two-socket AMD EPYC 7F52 16-Core Processor with hyper-threading running with 64 openMP threads compiled with gcc 11.4.0. In upper part of table the dependence on grid size in the radial variable $r_z$ is shown. We compare timings for 3 different modes of code running: solver without dual numbers (without AD), solver with two derivatives in $r$, solver with two derivatives in $r$, $Q_0$ and $\Lambda$. 
    The lower part of table shows the timings for different modes of running: for that we assume that we can have arbitrary number of parameters in the initial condition (we measured times up to 5).
    \label{tab. performance}}
    \end{center}
\end{table}

\section{Applications}
\label{sec. applications}

In this Section we provide two applications of the ideas described above where automatic differentiation provides substantial improvement in efficiency and precision.

\subsection{Fitting experimental data}
\label{sec. fitting}

The main advantage of the software discussed here is the possibility of analytic evaluation of derivatives of the scattering cross section with respect to the parameters of the initial condition. The latter are needed when a fit to experimentally measured cross section is performed. Fits using BK evolution equation to experimental data have already been performed in the past \cite{Gotsman:2002yy, Albacete:2009fh,Ducloue:2019jmy,Lappi:2013zma,Beuf:2020dxl} notably to measurements of DIS experiments conducted at the HERA accelerator. The measurements consist in tables of the reduced cross section
\begin{equation}
    \sigma_r(x,y,Q^2)=F_2(x,Q^2)-\frac{y^2}{1+(1-y)^2}F_L(x,Q^2),
    \label{eq. sigma r}
\end{equation}
where $y=Q^2/s x $ is elasticity and $\sqrt s$ is the collision energy. $F_L$ can be assumed to be small and neglected in the following, therefore $\sigma_r = \sigma_r(x,Q^2)$. $F_2$ is estimated for different values of Bjorken $x$ and momentum transfer $Q^2$. On the theoretical side $F_2$ can be expressed as a combination of cross-sections
\begin{equation}
F_2(x,Q^2)=F_T(x,Q^2)+F_L(x,Q^2)=\frac{Q^2}{4\pi^2\alpha_{em}}(\sigma_T^{\gamma^*p}+\sigma_L^{\gamma^*p}),
\end{equation}
where
\begin{equation}
    \sigma^{\gamma^*p}_{T,L}(x,Q^2)=\sum_f \int d^2 \brt \int_0^1 dz\, |\Psi^f_{T,L}(\brt,z,Q^2)|^2\sigma^{q\bar q}(\brt, x),
    \label{sig_LT_def}
\end{equation}
and the sum over $f$ usually extends over three lightest quarks which we assume to be massless.
$\Psi^f$ is the photon's wave-function and its interaction with quark flavor $f$ can be calculated in perturbation theory. At leading leading order in the QED coupling $\alpha_{em}$ (and zeroth order in $\alpha_s$), one has
\begin{equation}
    |\Psi_T^f(\brt, z,Q^2)|=\frac{3\alpha_{em}}{2\pi^2}e_{f}^2 \Big( z^2+(1-z)^2 \Big) \bar{Q}^2K_1^2(\bar{Q} |\brt|), \quad
    |\Psi_L^f(\brt,z,Q^2)|^2=\frac{3\alpha_{em}}{2\pi^2}e_{f}^2 4 Q^2 z^2 (1-z)^2K_0^2(\bar{Q} |\brt|),
\end{equation}
where $e_f$ is its electric charge, $K_0$ and $K_1$ are modified Bessel functions of the second kind, and 
\begin{equation}
    \bar{Q}^2=z(1-z)Q^2.
\end{equation}
The cross-section for the scattering of a pair of quark and anti-quark, $\sigma^{q\bar q}(\brt, x)$ is directly related to the dipole amplitude $S(\brt, \eta)$ discussed above via
\begin{equation}
    \sigma^{q\bar q}(\brt,x)=\sigma_0  \Big( 1 - S(\brt, \eta=\text{ln}(1/x)) \Big),
\end{equation}
where we used the large target approximation assuming that the proton target is uniform and infinite. $\sigma_0$ is a dimensionful parameter that has to be fixed through a comparison with experimental data. 

\begin{figure}
\begin{center}
\includegraphics[width=0.495\textwidth]{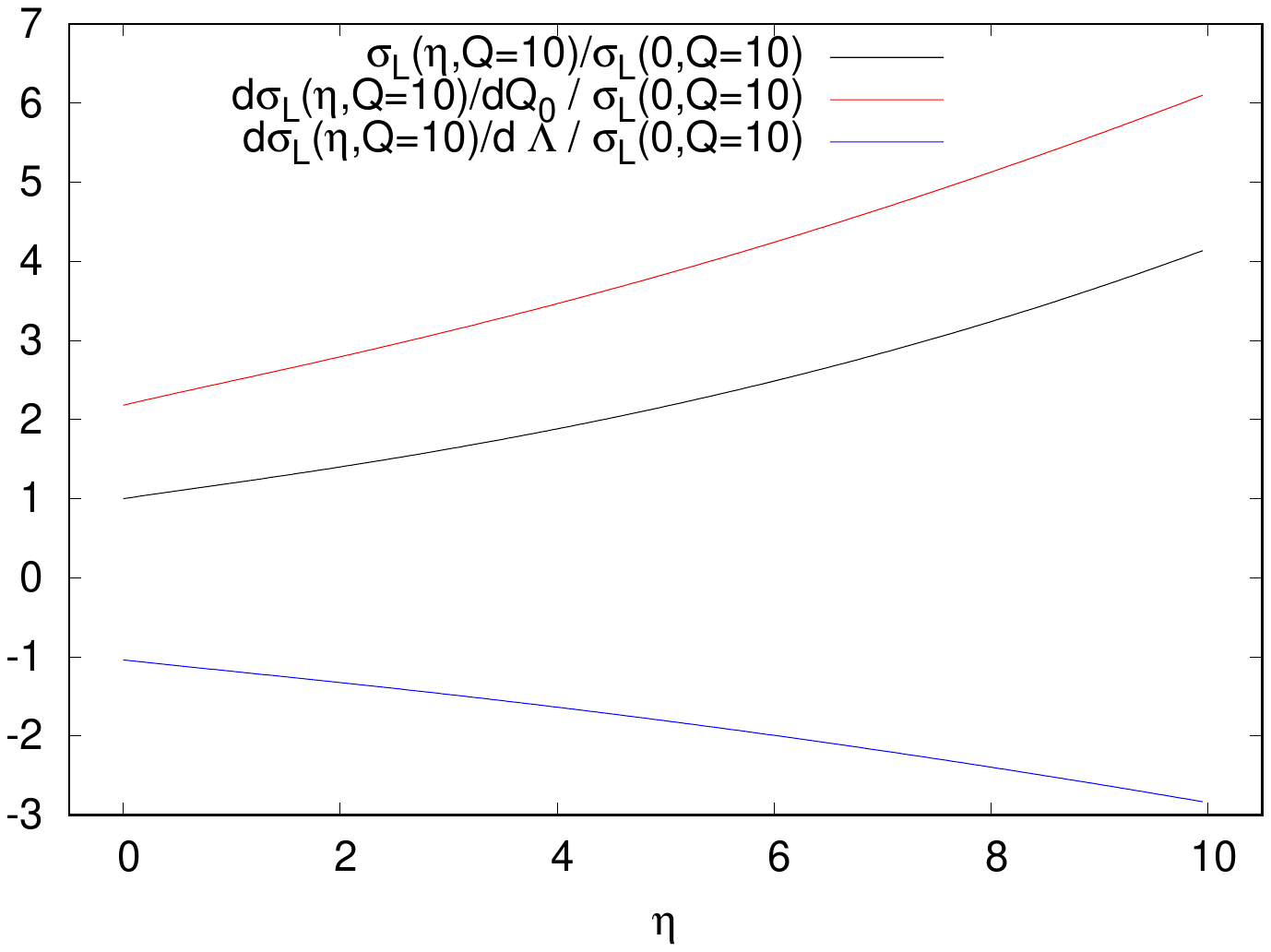}
\includegraphics[width=0.495\textwidth]{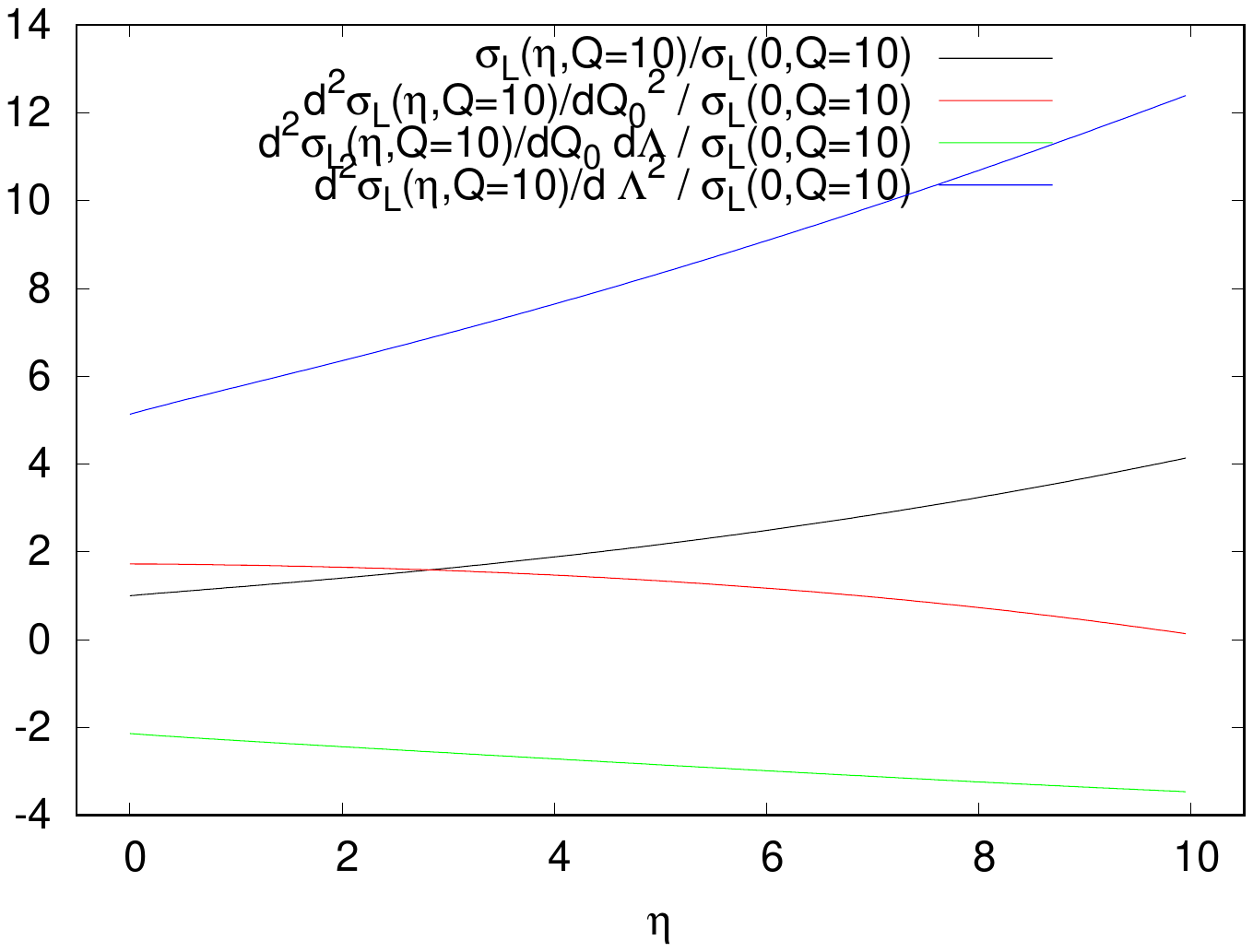}
\caption{First (left panel) and second (right panel) derivatives of the longitudinal cross-section $\sigma_L^{\gamma^*,p}(x,Q^2=10  \, \text{GeV}^2)$ given by Eq.~\eqref{sig_LT_def} with respect to the initial condition parameters. These derivatives can be directly used to estimate the gradient of the $\chi^2$ function during the optimization algorithm comparing theoretical predictions to the experimental measurements. \label{fig. cross section}}
\end{center}
\end{figure}

In the most naive approach, in order to perform the fit to experimental data one defines a $\chi^2$ function of the form
\begin{equation}
    \chi^2(\alpha,\beta,\dots) \propto \sum_{\textrm{experimental measurements}} \Big( \sigma_r(x,Q^2; \alpha, \beta, \dots) - \sigma^{\textrm{experimental}}_r(x,Q^2) \Big)^2 ,
    \label{eq. chi2}
\end{equation}
where $\alpha,\beta$ are the parameters of the initial condition and other free parameters introduced in the theoretical calculation of the reduced cross-section. We would like to find the optimal values that best describe the experimental measurements. In the optimization algorithm one requires the knowledge of the gradient of $\chi^2$ with respect to $\alpha, \beta$ in order to advance in the parameters space. The gradients of $\chi^2$ can be directly estimated from the gradients of $\sigma_r(x,Q^2; \alpha, \beta, \dots)$ which are given automatically by our software. In Fig. \ref{fig. cross section} we show the first and second derivatives of $\sigma^{\gamma^*p}_{L}(x,Q^2)$ cross-section which can be immediately translated into the gradients of $\chi^2$ given by Eq.~\eqref{eq. chi2} without the need to perform any additional simulation. We used the initial condition from Eq.~\eqref{MV_init_cond} with two differentiable parameters: $Q_0$ and $\Lambda$.

\subsection{Calculating Transverse Momentum Distributions (TMD)}
\label{sec. tmd}

In order to compute the cross section for some scattering processes at high energy, it is necessary to have the knowledge of transverse momentum dependent distributions (TMDs), which describe the distribution of partons within the hadron \cite{Boussarie:2023izj}. 
At high energy, it can be demonstrated that with a number of simplifications, 
the TMDs can be expressed as functions of the dipole distribution 
$S(r, \eta)$ and its derivatives {w.r.t.}~$r$, see \cite{Kharzeev:2003wz,Dominguez:2011br,Metz:2011wb,Lappi:2017skr,Boussarie:2021ybe} (some further details are provided  in \ref{tmd_appendix}). The rapidity dependence $\eta$ of the dipole distribution is obtained by solving the BK equation. In essence, the solution of the BK equation provides access to the function 
$S(r, \eta)$ and its associated derivatives. However, in practice, the value of 
$S(r, \eta)$ is only ever known at specific values of $r$ due to the discretization of the variable $r$ on the grid. Consequently, derivatives are only approximated using a number of points on the grid. For larger values of $r$ this may give rise to large discretization artefacts, as we will show below.

The AD method solves the above problem since $r$ can be a parameter over which the automatic differentiation is performed. As an example of application of our code we consider the quantities
\begin{align}
    \Kcal_\pm(r,\eta) = \frac{1}{\Gamma(r,\eta)}\left( \frac{\partial^2}{\partial r^2}  \pm \frac{1}{r} \frac{\partial}{\partial r} \right) \Gamma(r,\eta),
    \label{K_fuctions_def}
\end{align}
where 
\begin{align}
    \Gamma(r,\eta) = -\log\left[ S(r,\eta)\right].
\end{align}
It can be demonstrated that under a number of assumptions, namely when evaluating the TMDs in the high energy limit using the mean field and Gaussian approximations and treating the quadrupole correlation function as in \cite{Lappi:2017skr,Dominguez:2011wm,Dumitru:2011vk}, 
$\Kcal_{+}(r,\eta)$ ($\Kcal_{-}(r,\eta)$) appears in the expression for unpolarized (linearly polarized) gluon Weizsäcker-Williams (WW) TMDs. Such WW distributions are necessary ingredients to calculate the cross section for back-to-back production of inclusive dijets in DIS \cite{Dominguez:2011wm,Caucal:2022ulg,Caucal:2023fsf,Caucal:2023nci}. 

\begin{figure}
\begin{center}
\includegraphics[width=0.48\textwidth]{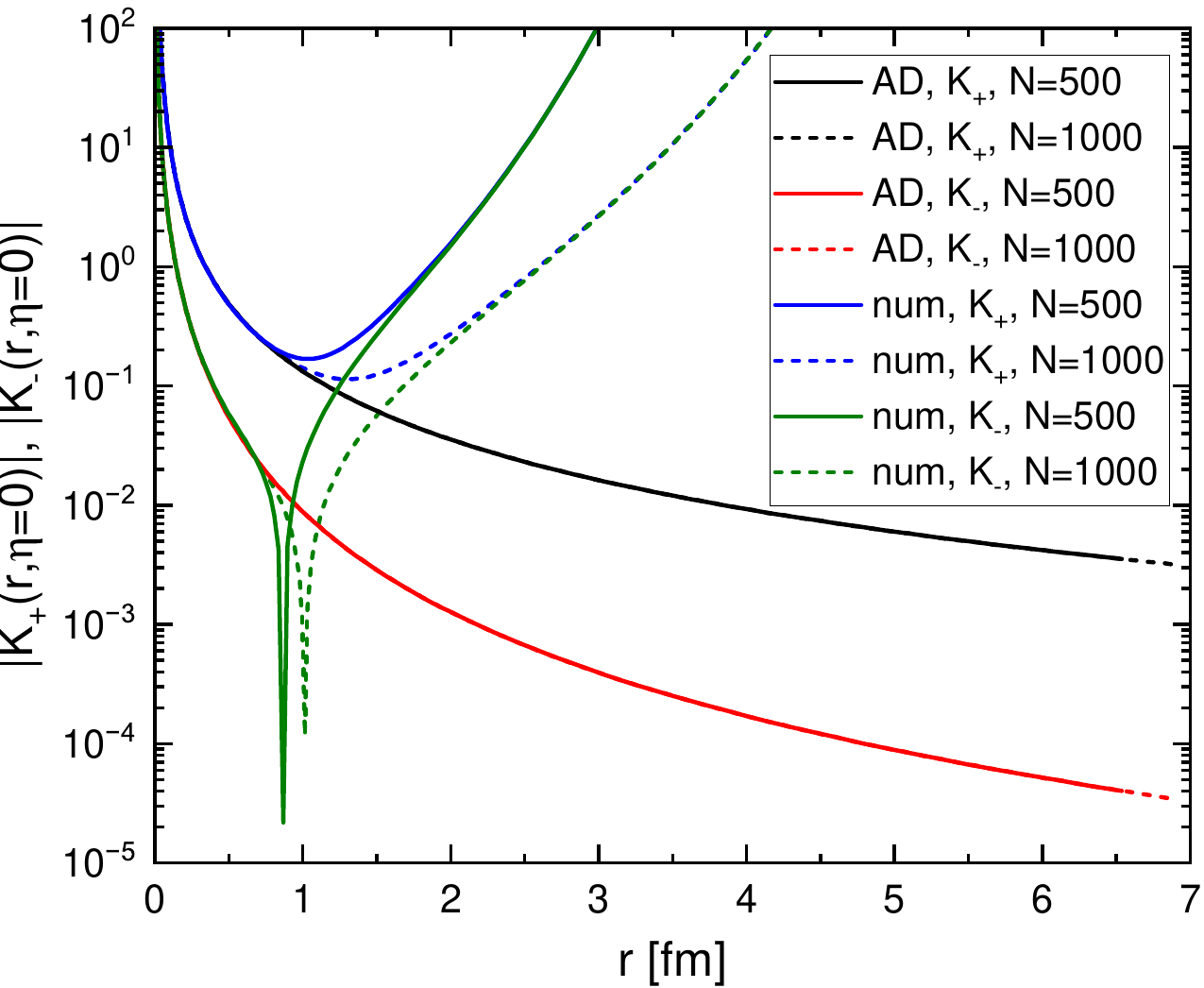}
\hfill
\includegraphics[width=0.48\textwidth]{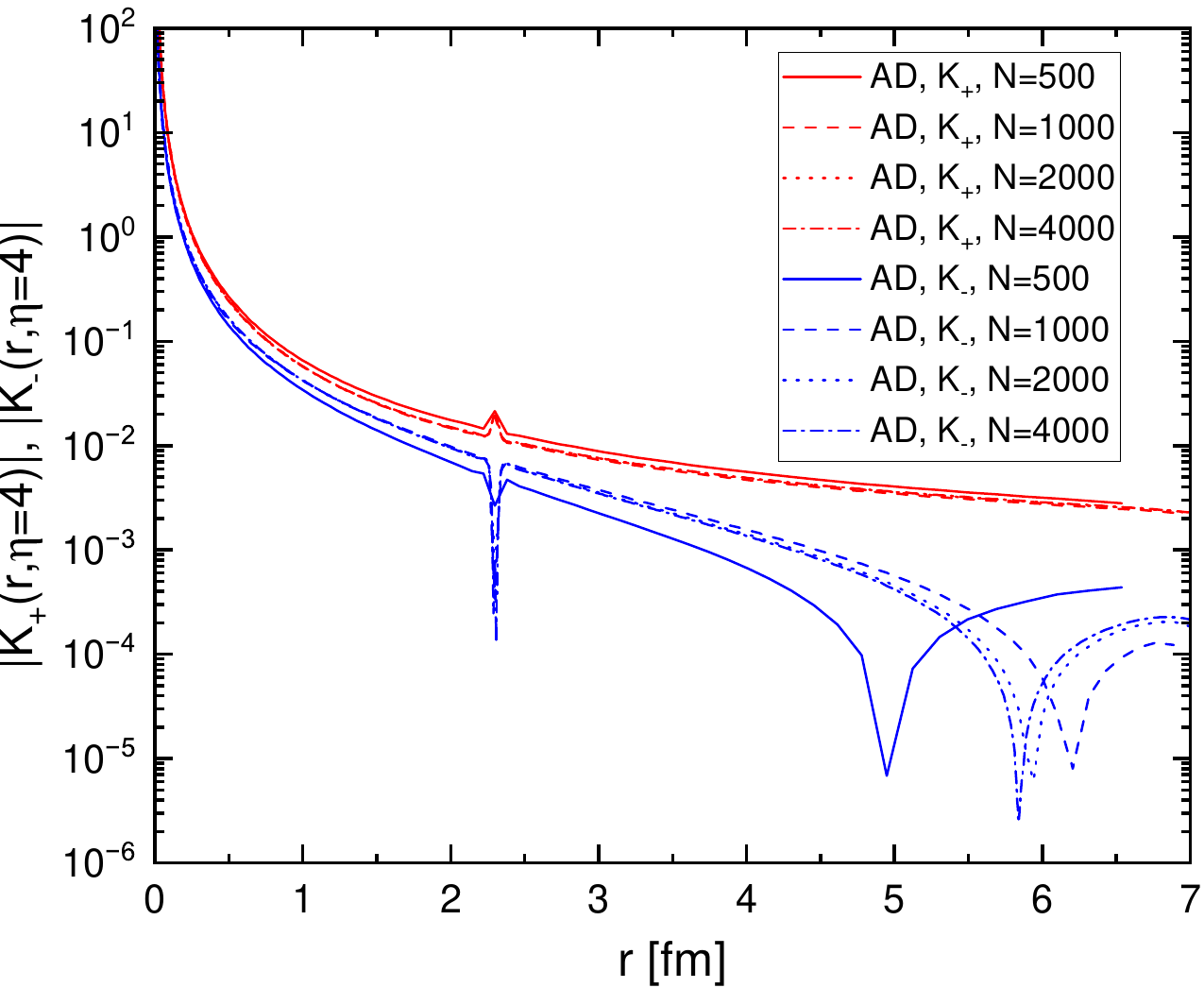}
\caption{Left panel: functions $|\Kcal_+(r_\perp)|$ and $|\Kcal_-(r_\perp)|$ at initial condition $\eta=0$ for two grid sizes, $N=500,\ 1000$ and two methods of evaluation: AD and divided difference on the grid points. Right panel: functions $|\Kcal_\pm(r_\perp)|$ after evolution to $\eta=4$ for four grid sizes: $N=500,\ 1000, \ 2000, \ 4000$.  }
\label{K_figures}
\end{center}
\end{figure}

It is worth noticing that although 
$S(r,\eta)$ usually quickly decreases with $r$, the
$\Kcal_\pm(r,\eta)$ decays much slower. For example, for the Gaussian form of the dipole amplitude  
$S(r,\eta)= \exp(-r^2 Q_{0}^2)$ one gets $\Kcal_{+}(r,\eta)=4 r^{-2}$. 
As $\Kcal_{+}(r,\eta)$ and $\Kcal_{-}(r,\eta)$ appear in convolutions over $r$, good control of numerical precision at large $r$ is necessary.
In Figure \ref{K_figures} we show the result for the absolute values of 
$\Kcal_{+}(r,\eta)$ and $\Kcal_{-}(r,\eta)$ calculated using our software. The left panel shows the functions for the initial condition, so only derivatives of Eq.~\eqref{MV_init_cond} are involved. We compare the AD results with the numerical estimate of the derivatives calculated from neighboring points on a $r$ grid. Solid (dashed) lines denote results obtained using a grid with 500 points (1000 points) in the $r$ variable. 
It is clear that numerical derivatives (red and green curves) give unreliable results for large values of $r$. This is due to the fact that the logarithmic grid becomes sparser at large values, thus increasing the numerical error of the divided difference. 
Comparing results from 500 and 1000 point grids, one sees the significant difference between those two. Automatic differentiation, on the other hand, is not sensitive to mesh size because it provides analytically computed values of derivatives for each node on the grid. This can be seen from the blue and yellow lines, where plain ($500$ points) and dashed ($1000$ points) distributions converge to the same values.

The right panel of Figure \ref{K_figures} shows absolute values of $\Kcal_\pm$ functions after evolution to $\eta=4$ obtained using AD. Different curve styles indicate the grid sizes used. One observes some dependence of grid size at large $r$. This is due to the fact that during the evolution one accumulates discretization errors from the evaluation of the integral in Eq.~\eqref{eq. BK improved}. We note that for $N=2000$ and $N=4000$ the results are very close to each other. We noticed some numerical artifact around $r\approx r_{max} \approx 2.5\, \text{fm}$, which is the point where $\bar{\alpha}_s$ is frozen (see Eq.~\eqref{eq. alpha_s}) and where the first derivative is not smooth.

\section{Conclusions}
\label{sec. conclusions}

In this work, we have demonstrated that integrating the technique of automatic differentiation into the solver of the Balitsky-Kovchegov evolution equation can yield multiple advantages. On the one hand, the knowledge of the gradients and the hessian matrix with respect to the parameters of the initial condition can simplify and speed up the fitting procedure. This should facilitate and automate the investigation of more complicated forms of initial conditions. On the other hand, we have shown that automatic differentiation applied to the variable $r$, which is the separation between quark and anti-quark in the dipole, allows to study transverse momentum dependent parton distributions at much larger distances. Thanks to the fact that automatic differentiation is able to evaluate the derivatives analytically, the behavior of $\mathcal{K}_-$ and $\mathcal{K}_+$ at large $r$ is numerically stable. In addition, what has not been mentioned so far, but should be clear from the above discussion, is that one can use automatic differentiation on the various parameters introduced in the algorithm, such as the regularization of the dipole kernel, the value at which the current coupling is frozen, etc. By promoting these parameters to dual numbers, the simulation would automatically yield the dependence of the final results on the values of these parameters, thus helping to study the sensitivity of the results to these parameters. A similar extension may be applicable to the solution of the more general JIMWLK evolution equation. The JIMWLK equation is solved for for individual Wilson lines, so the dipole size is defined only for correlations of Wilson lines.
However, the dependence of the final results on the parameters of the initial condition is of great importance and the gain in performance offered by automatic differentiation may be even greater due to the higher numerical cost of solving the JIMWLK equation.

\section*{Acknowledgements}
We thank Farid Salazar and Wenbin Zhao for discussions that gave us inspiration for this work. We thank Paul Caucal for sharing his code with us and help with running it. We thank Piotr Białas for his comments on the draft.
We also thank Leszek Motyka for discussions on the BK equation and
Alberto Ramos for discussions on automatic differentiation.

We gratefully acknowledge Polish high-performance computing infrastructure PLGrid (HPC Center: ACK Cyfronet AGH) for providing computer facilities and support within computational grants no. PLG/2022/015321 and no. PLG/2023/016656. T.S. kindly acknowledges the support of the Polish National Science Center (NCN) grant No. 2021/43/D/ST2/03375. P.K. and F.C. acknowledge support from of the Polish National Science Center (NCN) grant No. 2022/46/E/ST2/00346.

\appendix

\section{Implementation details}
{\label{sec. implementation}}

In this Section, we provide some additional, and technical details concerning the implementation.

In the listing \ref{list. dual class} we show the class which implements the dual numbers. It contains one parameter adjustable by the user at compile time, namely $NN$ that defines the number of differentiable parameters to be included in the calculation. This number does not include the radial variable which is kept separate. The class contains containers where the first and second derivatives are stored and accumulated.

\begin{lstlisting}[language=C++, caption=Dual number class called \emph{dual} set up for the calculation of derivatives with respect to two parameters and the dipole size., label=list. dual class]
//number of parameters with derivatives
const int NN = 2;
//number of independent entries in the hessian matrix
const int NN_hessian = NN*NN - NN*(NN-1)/2;

template <typename T> class dual {
        public:
        T val;
        T derr;
        T der2r;
        T der[NN];
        T der2[NN_hessian];

        dual(T v){
                val = v;
                derr = 0.0;
                der2r = 0.0;
                for(int i=0; i<NN; i++)
                        der[i] = 0.0;
                for(int i=0; i<NN_hessian; i++)
                        der2[i] = 0.0;
        }
};
\end{lstlisting}

Listing \ref{list. mult operator} shows the overloading of the multiplication operator for dual numbers. Infinitesimal arithmetics for the first and second derivatives are implemented separately for the radial variable and for the set of differentiable parameters. Similar overloading is done for all other basic operations appearing in the algorithm. Additional functions are added to treat mixed-type operations, for instance, the multiplication of a dual number by a real number.

\begin{lstlisting}[language=C++, caption=Overloading of the multiplication operator for dual numbers., label=list. mult operator]
template <typename T> dual<T> operator*(dual<T> a, dual<T> b){

        dual<T> c;
        c.val = a.val*b.val;
        c.derr = a.derr*b.val + a.val*b.derr;
        c.der2r = a.der2r*b.val + 2.0*a.derr*b.derr + a.val*b.der2r;
        for(int i=0; i<NN; i++)
                c.der[i] = a.der[i]*b.val + a.val*b.der[i];
        for(int i=0; i<NN; i++)
                for(int j=i; j<NN; j++){
                        int index = i*NN+j-i*(i+1)/2;
                        c.der2[index] = a.der2[index]*b.val 
                        + a.der[i]*b.der[j] + a.der[j]*b.der[i] 
                        + a.val*b.der2[index];
                }
        return c;
}
\end{lstlisting}

Listing \ref{list. exp function} contains a wrapper of the exponential function. Apart from the single call to the standard library $\exp$ function, the calculation of the derivatives is implemented.

\begin{lstlisting}[language=C++, caption=Wrapper of the $\exp$ function. \label{list. exp function}]
template <typename T> dual<T> expp(dual<T> x){

        dual<T> z;
        z.val = exp(x.val);

        z.derr = z.val*x.derr;
        z.der2r = x.derr*z.derr + z.val*x.der2r;
        for(int i=0; i<NN; i++)
                z.der[i] = x.der[i]*z.val;
        for(int i=0; i<NN; i++)
                for(int j=i; j<NN; j++){
                        int index = i*NN+j-i*(i+1)/2;
                        z.der2[index] = z.der[i]*x.der[j] + z.val*x.der2[index];
                }
        return z;
}

\end{lstlisting}

\section{TMD in Color Glass Condensate}
\label{tmd_appendix}
In high-energy limit, strong intersections can be described within effective field theory called Color Glass Condensate (CGC) \cite{Iancu:2003xm,Weigert:2005us,Jalilian-Marian:2005ccm,Gelis:2010nm,Albacete:2014fwa,Kovchegov:2012mbw,Morreale:2021pnn}. In this picture, the parton (gluon or quark) with high longitudinal momentum scatters off the classical gluonic field $A^{a}_{\rm cl}(z^-,\xt)$. Degrees of freedom are given by Wilson lines:
\begin{equation}
    V(\xt)=\mathcal{P}\exp\left(ig\int_{-\infty}^{\infty}\der z^-A^{+,a}_{\rm cl}(z^-,\xt)\right)\,.
\end{equation}
where only $+$ component of the field contributes. 

In CGC one can factorize the cross-section for the scattering into perturbative hard factors and non-perturbative correlators of the Wilson lines. In particular, at high energy one can find the connections between the transverse momentum distributions (TMDs) and the Wilson lines. As an example, lets consider Weizsäcker-Williams TMD  which is given by:
\begin{align}
    G^{ij}_{WW}(\kt,x=e^{-\eta}  )&=\int\frac{\der^2\bt\der^2\bt'}{(2\pi)^4}e^{-i\kt\cdot(\bt-\bt')} \frac{-2}{\alpha_s}\left\langle\Tr\left[V(\bt) \left(\partial^iV^\dagger(\bt) \right) V(\bt') \left(\partial^jV^\dagger(\bt') \right)\right]\right\rangle_{\eta}\,, 
    \label{gluon_WW_TMD}
\end{align}
where the average denoted by $\left\langle \cdots \right\rangle_{\eta}$ is performed over classical color field configurations. At this point evaluating WW TMD requires solution of JIMWLK equation \cite{Jalilian-Marian:1997jhx,Jalilian-Marian:1997ubg,Kovner:2000pt,Iancu:2000hn,Ferreiro:2001qy,Iancu:2001ad,Weigert:2000gi} which allows to calculate arbitrary correlators of Wilson lines. However, solving JIMWLK is  numerically expensive and no implementation of the kinematical constraint is currently available \cite{Korcyl:2024xph,Korcyl:2024zrf}. To overcome this issue, one applies large $N_c$ and Gaussian approximations \cite{Fujii:2006ab,Marquet:2010cf} which allow to express arbitrary correlator as a function of just one object, namely two-point correlation function: 
\begin{align}
    S(r, \eta) = \frac{1}{N_c} \left \langle \Tr\left[ V(\xt) V^\dagger(\yt) \right] \right \rangle_\eta.
\end{align}
These approximations were checked to work well for leading-logarithmic evolution  \cite{Dumitru:2011vk}. Applying them to (\ref{gluon_WW_TMD}) one gets:
\begin{align}
        \alpha_s \Fcal_{WW}(\kt,x=e^{-\eta}) &\equiv \delta_{ij} G^{ij}_{WW}(\kt,x=e^{-\eta}  )= \frac{C_F S_\perp }{2\pi^2} \int \frac{r\, \der r}{2\pi} J_0(k_\perp r) \Kcal_+(r,\eta)\left[1 - \left( S(r,\eta) \right)^{N_c/C_F} \right] \label{eq:FWW}
\end{align}
for unpolarized TMD and
\begin{align}
    \alpha_s \Hcal_{WW}(\kt,x=e^{-\eta}) &\equiv  \left(\frac{2\kt^i \kt^j}{\kt^2} -\delta^{ij}\right)G^{ij}_{WW}(\kt,x=e^{-\eta}) = -\frac{C_F S_\perp }{2\pi^2} \int \frac{r\, \der r}{2\pi} J_2(k_\perp r) \Kcal_-(r,\eta)\left[1 - \left( S(r,\eta) \right)^{N_c/C_F} \right] \label{eq:FWW2}
\end{align}
for linearly polarized TMD. In above expressions $C_F=4/3$, $S_\perp$ is transverse size of the hadron, $J_0$ and $J_2$ are Bessel functions of the first kind, and $\Kcal_\pm$ are defined in eq. (\ref{K_fuctions_def}).

\bibliography{ref}
\end{document}